\documentclass{iau}
\usepackage{graphicx}
\usepackage{soul,color,natbib}

\title[The effects of stellar winds and magnetic fields on exoplanets] 
{The effects of stellar winds and magnetic fields on exoplanets}

\author[A.~A.~Vidotto] {A.~A.~Vidotto}

\affiliation{SUPA, University of St Andrews, North Haugh, KY16 9SS, UK \\email: {\tt Aline.Vidotto@st-andrews.ac.uk} }

\pubyear{2013}
\volume{302}  
\pagerange{}
\jname{Magnetic fields throughout stellar evolution}
\editors{P. Petit, eds.}

\begin{document}
\maketitle

\begin{abstract}
The great majority of exoplanets discovered so far are orbiting cool, low-mass stars whose properties are relatively similar to the Sun. However, the stellar magnetism of these stars can be significantly different from the solar one, both in topology and intensity. In addition, due to the present-day technology used in exoplanetary searches, most of the currently known exoplanets are found orbiting at extremely close distances to their host stars ($< 0.1$ au). The dramatic differences in stellar magnetism and orbital radius can make the interplanetary medium of exoplanetary systems remarkably distinct from that of the Solar System. To constrain interactions between exoplanets and their host-star's magnetised winds and to characterise the interplanetary medium that surrounds exoplanets, more realistic stellar wind models, which account for factors such as stellar rotation and the complex stellar magnetic field configurations of cool stars, must be employed. Here, I briefly review the latest progress made in data-driven modelling of magnetised stellar winds. I also show that the interaction of the stellar winds with exoplanets can lead to several observable signatures, some of which that are absent in our own Solar System.
\keywords{MHD, stars: magnetic fields, stars: mass loss, stars: planetary systems, stars: rotation, stars: winds, outflows}
\end{abstract}

\firstsection 
\section{Introduction}

Stars experience mass loss in the form of winds during their lives and, as the star evolves, its wind changes characteristics. For some stars, at certain evolutionary phases, the amount of mass lost through stellar winds can amount to a significant portion of its own mass. This is the case, for example, of supergiant stars, which present massive winds that are believed to be mainly driven by radiation pressure on dust grains (for the red supergiants) or on spectral lines (for the blue supergiants). On the other hand, for cool stars at the main sequence phase, we believe their winds are significantly less massive, carrying only a small amount of mass compared to the total mass of the star. Although these winds are quite rarefied, they carry a significant amount of angular momentum, which affects the stellar rotational evolution \citep[e.g.,][]{1997A&A...326.1023B}. 

P-Cygni profiles, the traditional mass-loss signatures observed in the denser winds of giant and supergiant stars, are not formed in the rarefied winds of cool dwarf stars. To detect these winds, other indirect methods have been proposed \citep[e.g., ][]{1996ApJ...462L..91L,2001ApJ...547L..49W, 2002ApJ...578..503W}. So far, the method developed by \citet{2001ApJ...547L..49W} has been the most successful one, enabling estimates of mass-loss rates for about a dozen cool dwarf stars. 

Due to the lack of observational constraints on winds of cool low-mass stars, many works assume  these winds to be identical to (or a scaled version of) the solar wind. In the next Section, I highlight a few observed properties of cool dwarf stars that suggest that their winds can actually be significantly different from the solar wind.

\subsection{What can we infer from winds of cool stars?}

\subsubsection{Temperatures}
We believe the solar coronal heating is ultimately caused by the Sun's magnetism, although it is still debatable which mechanism heats the solar corona. In the stellar case, coronal heating is much less understood. Stellar coronae can be much hotter than the solar corona. The temperature of the X-ray emitting coronae of solar analogs can exceed $10$~MK \citep{2004A&ARv..12...71G}, more than one order of magnitude larger than the solar coronal temperature of $\sim 1$~MK. 
If the temperature of the X-ray emitting (closed) corona is related to the temperature of the stellar wind (flowing along open field lines), as one would naively expect, then we may expect stellar winds of cool dwarf stars to have temperatures that could be much larger than the solar wind temperature. 
Because the terminal velocities of thermally-driven winds are very sensitive to the choice of temperature of the stellar wind \citep{1958ApJ...128..664P}, the winds of cool dwarf stars might have a variety of terminal velocities.

\subsubsection{Magnetism}\label{sec.magnetism}
Thanks to our privileged position immersed in the solar wind, we have access to a great quantity of data that allow a detailed understanding of the physics that is operating in the Sun. Measurements of mass flows and magnetic field of the solar wind have revealed an asymmetric solar wind \citep{1995JGR...10019893M, 1996GeoRL..23.3267S, 1998GeoRL..25.3109J, 2006A&A...455..697W}. In particular, Ulysses measurements showed that the solar wind structure depends on the characteristics of the solar magnetic field \citep{2008GeoRL..3518103M}. The solar wind characteristics change along the solar cycle, presenting a simple bimodal structure of fast and slow flows during solar minimum, when the geometry of the solar magnetic field is closest to that of an aligned dipole. At solar maximum, when the the axis of the large-scale dipole becomes nearly perpendicular to the solar axis of rotation, the solar wind shows a more complex structure.

Although the richness of details of the magnetic field configuration is only known for our closest star, modern techniques have made it possible to reconstruct the large-scale surface magnetic fields of other stars. The Zeeman-Doppler Imaging (ZDI) technique is a tomographic imaging technique \citep{1997A&A...326.1135D} that allows us to reconstruct the large-scale magnetic field (intensity and orientation) at the surface of the star from a series of circular polarisation spectra. This method has now been used to investigate the magnetic topology of stars of different spectral types \citep{2009ARA&A..47..333D} and has revealed fascinating differences between the magnetic fields of different stars.  For example, solar-type stars that rotate about two times faster than our Sun show the presence of substantial toroidal component of magnetic field, a component that is almost non-existent in the solar magnetic field \citep{2008MNRAS.388...80P}. The magnetic topology of low-mass ($<0.5~M_\odot$) very active stars seem to be dictated by interior structure changes: while partly convective stars possess a weak non-axisymmetric field with a significant toroidal component, fully convective ones exhibit strong poloidal axisymmetric dipole-like topologies \citep{2008MNRAS.390..567M,2008MNRAS.390..545D}.

\subsubsection{Rotation}
In addition to magnetic field characteristics, the symmetry of a stellar wind also depends on the rotation rate of the star. In the presence of fast rotation, a magnetised wind can lose spherical symmetry, as centrifugal forces become more important with the increase of rotation rate \citep{1993MNRAS.262..936W}. The distribution of rotation periods in cool main-sequence stars is very broad, ranging from stars rotating faster than once per day to indefinitely long periods. 

\subsection{Summary}
The variety of observed rotation rates, intensities and topologies of the magnetic fields of cool, dwarf stars indicate that their winds might come in different flavours and might be significantly different from the solar one. One might also bear in mind that, similarly to the Sun, low-mass stars are also believed to host magnetic and activity cycles (see Fares contribution, this volume), which imply that the characteristics of their winds can also vary in a time scale of the cycle periods.

\section{Models and simulations of winds of cool stars}
There are two most commonly used ways to model stellar winds, which I summarise below. 
\begin{enumerate}
\item One approach consists of computing the detailed energetics of the wind, starting from the photosphere, passing through the chromosphere, until it reaches the stellar corona \citep[e.g.,][]{1973ApJ...181..547H, 1983ApJ...275..808H, 1980ApJ...242..260H, 1989A&A...209..327J, 2006ApJ...639..416V, 2006MNRAS.368.1145F, 2008ApJ...689..316C, 2011ApJ...741...54C, 2012arXiv1212.6713S}. Depending on the physics that is included in such models, this approach can become computationally expensive. As a result, it has been limited to analytical, one- and two-dimensional solutions. In addition, it is also usually focused in the inner most part of the corona and usually only adopts simplified magnetic field topologies. 
\item The second approach commonly used to model winds of cool stars consists of adopting a simplified energy equation, usually assuming the wind to be isothermal or described by a polytropic equation of state (in which thermal pressure $p$ is related to density $\rho$ as  $p \propto \rho^\gamma$). In this case, one is allowed to perform multi-dimensional numerical simulations of stellar winds and can incorporate more complex magnetic field topologies \cite[e.g.,][]{1968MNRAS.138..359M, 1971SoPh...18..258P,1989ApJ...342.1028T, 1993MNRAS.262..936W, 2000ApJ...530.1036K,2001A&A...371..240L, 2009ApJ...699..441V,2009ApJ...703.1734V,2010ApJ...720.1262V, 2011ApJ...737...72P, 2013MNRAS.431..528J}
\end{enumerate}

Because of the simplified energetics that are considered in the second approach, its domain can extend considerably farther out than the first one. As a result, polytropic winds can be useful in the characterisation of the interplanetary medium and also to characterise interactions between exoplanets and the winds of their host-stars. In the next Section, I present how to make more realistic stellar wind simulations by incorporating recent insights acquired on the magnetic topology of different stars into stellar wind models. 

\subsection{Data-driven wind simulations}
To illustrate how observationally reconstructed surface maps (see Section \ref{sec.magnetism}) can be incorporated in the simulations of stellar winds, I will present the work done in \cite{2012MNRAS.423.3285V}, where we performed numerical simulations of the stellar wind of the planet-hosting star $\tau$~Boo (spectral type F7V). $\tau$~Boo is a remarkable object, not only because it hosts a giant planet orbiting very close to the star (located at 0.046~au from the star), but also because it is the only star other than the Sun for which a full magnetic cycle has been reported in the literature \citep{2008MNRAS.385.1179D, 2009MNRAS.398.1383F,2013arXiv1307.6091F}. These observations suggest that $\tau$~Boo undergoes magnetic cycles similar to the Sun, but with a cycle period that is about one order of magnitude smaller than the solar one (about 2 years as opposed to 22 years for the solar magnetic cycle). 

\begin{figure}
\centering
	\includegraphics[width=100mm]{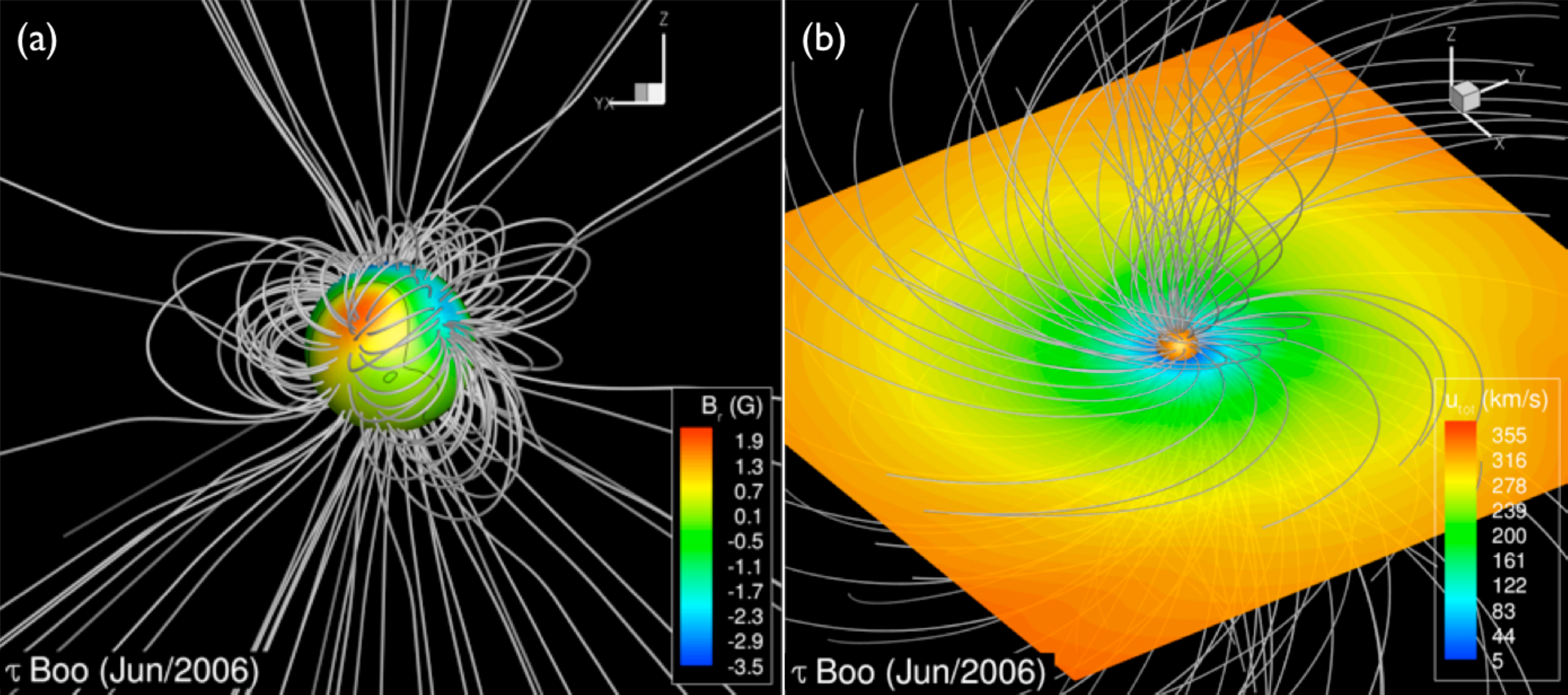}
\caption{(a) Potential field extrapolation that is used as initial condition for the stellar wind simulations. The distribution of the observationally reconstructed surface magnetic field (used as boundary condition) is shown colour-coded. (b) Self-consistent solution of the magnetised stellar wind after the solution relaxed in the grid. The wind velocity is shown at the equatorial plane. Figure adapted from \cite{2013A&G....54a1.25V}.}\label{fig.sim_conditions}
\end{figure}

The surface magnetic maps of  $\tau$~Boo reconstructed by \cite{2007MNRAS.374L..42C}, \cite{2008MNRAS.385.1179D} and \cite{2009MNRAS.398.1383F} were used as boundary conditions for the stellar wind simulations. A potential field extrapolation is adopted at the initial state of the simulation (Fig.~\ref{fig.sim_conditions}a). As the simulation evolves in time, stellar wind particles and magnetic field lines are allowed to interact with each other. Figure \ref{fig.sim_conditions}b shows the self-consistent solution found for the magnetic field lines, after the wind solution relaxed in the grid. Note that the magnetic field lines become stressed, wraping around the rotational axis of the star (pointing towards positive $z$). Colour-coded are the reconstructed large-scale surface field of $\tau$~Boo at Jun/2006 (Fig.~\ref{fig.sim_conditions}a) and the wind velocity at the equatorial plane of the star (Fig.~\ref{fig.sim_conditions}b). 

\cite{2012MNRAS.423.3285V} found that variations of the stellar magnetic field during the cycle directly influence the outflowing wind. Therefore, the rapid variation of the large-scale magnetic field of $\tau$~Boo implies that the environment surrounding the close-in planet should be varying quite rapidly. In addition, \cite{2012MNRAS.423.3285V} estimated the mass-loss rate ($\dot{M}$) of $\tau$~Boo, showing that this star seems to have a denser wind than that of the Sun, with $\dot{M}$ that are 2 orders of magnitude larger than the solar value $\dot{M}_\odot$ ($\dot{M} \approx 135~\dot{M}_\odot$). 

\section{Interaction between stellar winds and exoplanets}
When the wind outflows from the star, it permeates the entire extrasolar system, interacting with any body that it encounters on its way. The interaction between stellar winds and exoplanets can lead to observable signatures, some of which are absent in our own solar system.

\subsection{Planetary radio emission}
In the solar system, the giant planets and the Earth emit at radio wavelengths. Such planetary radio emission is due to the interaction between the magnetic planets and the solar wind, and its power is proportional to the stellar wind energy dissipated in the wind-planet interaction. By analogy to what is observed in the solar system, it is expected that exoplanets also interact with the winds of their host stars and should, therefore, also generate radio emission. Because the energy dissipated by the stellar wind is larger at closer distances to the star (because the wind density and magnetic fields are significantly larger than further out from the star), radio emission of close-in planets, such as $\tau$~Boo~b, is expected to be several orders of magnitude larger than the emission from the planets in the solar system \citep{2007P&SS...55..598Z}. In addition, the relatively dense wind of $\tau$~Boo estimated in \cite{2012MNRAS.423.3285V} implies that the energy dissipated in the stellar wind-planet interaction can be significantly higher than the values derived in the solar system. Combined with the close proximity of the system ($\sim16$~pc), the $\tau$~Boo system has been one of the strongest candidates to verify exoplanetary radio emission predictions. 

Using the detailed stellar wind model developed for its host-star, \cite{2012MNRAS.423.3285V} estimated radio emission from $\tau$~Boo~b, exploring different values for the {\it assumed} planetary magnetic field. For example, they showed that, for a planet with a magnetic field similar to Jupiter's ($\simeq 14$~G), the radio flux is estimated to be $ \simeq 0.5-1$~mJy, occurring at an emission frequency of $\simeq 34$~MHz. Although small, this emission frequency lies in the observable range of current instruments, such as LOFAR. To observe such a small flux, an instrument with a sensitivity lying at a mJy level is required. The same estimate was done considering the planet has a magnetic field similar to the Earth ($\simeq 1$~G). Although the radio flux does not present a significant difference to what was found for the previous case, the emission frequency ($\simeq 2$~MHz) falls at a range below the ionospheric cut-off, preventing its possible detection from the ground. In fact, due to the ionospheric cutoff at $\sim 10~$MHz, radio detection  with ground-based observations from planets with magnetic field intensities  $\lesssim 4$~G should not be possible \citep{2012MNRAS.423.3285V}. 

\subsection{Bow shock signatures in transit observations}
Despite many attempts, radio emission from exoplanets has not been detected so far \citep[e.g.,][]{2009MNRAS.395..335S,2010AJ....139...96L,2013A&A...552A..65L}. Such a detection would not only constrain local characteristics of the stellar wind, but would also demonstrate that exoplanets are magnetised. Fortunately, there may be other ways to probe exoplanetary magnetic fields, in particular for transiting systems, through signatures of bow shocks during transit observations.

Based on Hubble Space Telescope/Cosmic Origins Spectrograph (HST/COS) observations using the narrow-band near-UV spectroscopy, \cite{2010ApJ...714L.222F} showed that the transit lightcurve of the close-in giant planet WASP-12b presents both an early ingress when compared to its optical transit, as well as excess absorption during the transit, indicating the presence of an asymmetric distribution of material surrounding the planet. Motivated by these transit observations, \cite{2010ApJ...722L.168V} suggested that the presence of bow shocks surrounding close-in planets might lead to transit asymmetries at certain wavelengths, such as the one observed in WASP-12b at near-UV wavelengths.   

The main difference between bow shocks formed around exoplanets and the ones formed around planets in the solar system is the shock orientation, determined by the net velocity of the particles impacting on the planet's magnetosphere. In the case of the Earth, the solar wind has essentially only a radial component, which is much larger than the orbital velocity of the Earth. Because of that, the bow shock surrounding the Earth's magnetosphere forms facing the Sun. However, for close-in exoplanets that possess high orbital velocities and are frequently located at regions where the host star's wind velocity is comparatively much smaller, a shock may develop ahead of the planet. In general, we expect that shocks are formed at intermediate angles \citep[see also][]{2011MNRAS.414.1573V, 2013arXiv1309.2938L}.

\begin{figure}
\centering
	\includegraphics[width=0.9\textwidth]{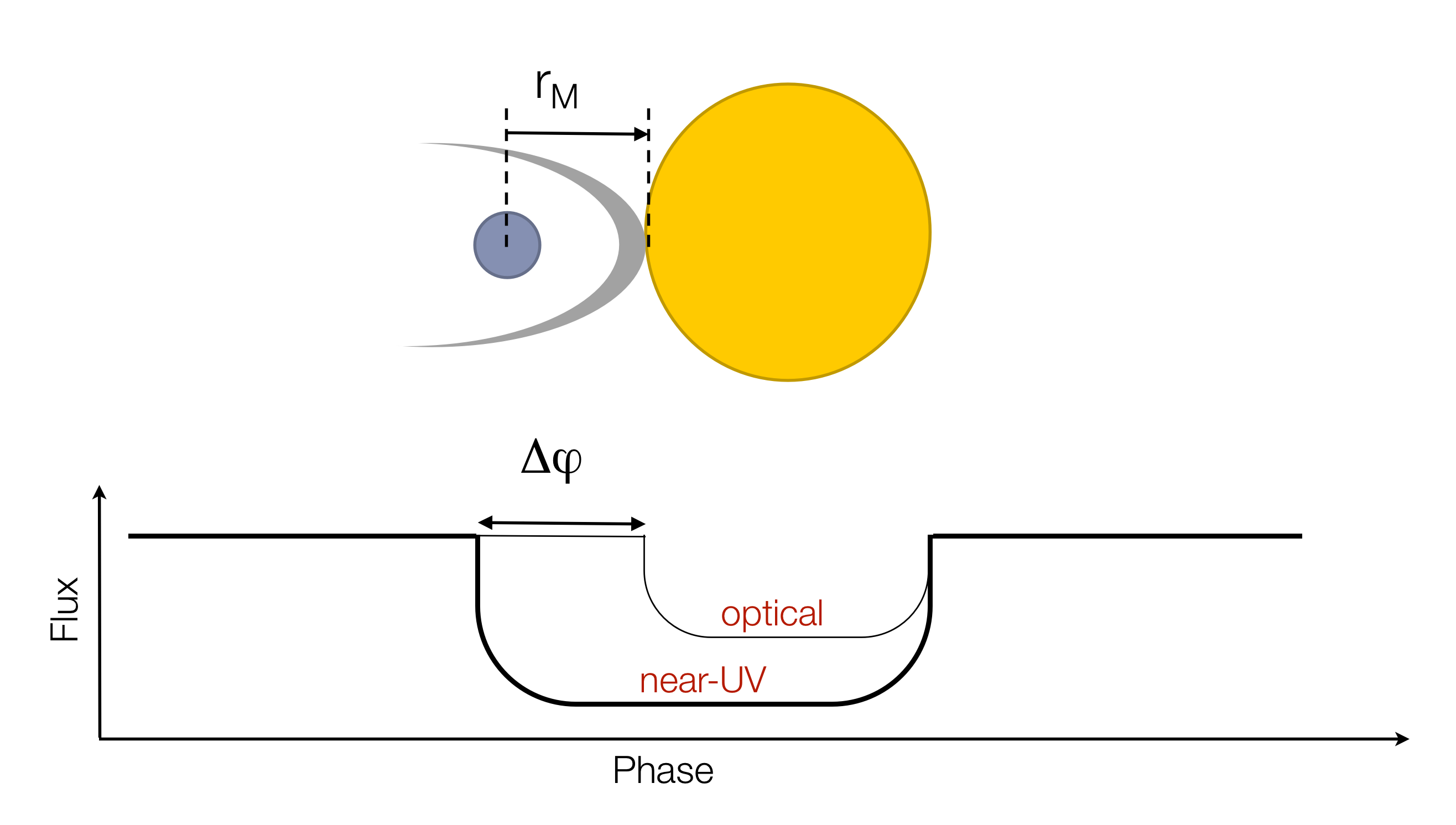}
\caption{Sketch of the light curves obtained through observations in the optical and in certain near-UV lines, where the bow shock surrounding the planet's magnetosphere is also able to absorb stellar radiation. The stand-off distance from the shock to the centre of the planet is assumed to trace the extent of the planetary magnetosphere $r_M$, which can be estimated by measuring the difference between the phases ($\Delta \varphi$) at which the near-UV and the optical transits begin. }\label{fig.sketch2}
\end{figure}

Due to their high orbital velocities, close-in planets offer the best conditions for transit observations of bow shocks. If the compressed shocked material is able to absorb stellar radiation, then the signature of bow shocks may be observed through both a deeper transit and an early-ingress in some spectral lines with respect to the broadband optical ingress \citep{2010ApJ...722L.168V}. The sketch shown in Figure \ref{fig.sketch2} illustrates this idea. 

In the case of WASP-12b, \cite{2010ApJ...722L.168V} suggested that the shocked material, which is able to absorb enough stellar radiation in the near-UV, can cause an asymmetry in the lightcurve observed (see lightcurve sketches in Figure~\ref{fig.sketch2}), where the presence of compressed material ahead of the planetary orbit causes an early ingress, while the lack of compressed material behind the planetary orbit causes simultaneous egresses both in the near-UV transit as well as in the optical one. This suggestion was verified by \cite{2011MNRAS.416L..41L}, who performed Monte Carlo radiation transfer simulations of the near-UV transit of WASP-12b. They confirmed that the presence of a bow shock indeed breaks the symmetry of the transit lightcurve

\subsection{Planetary magnetic fields: a new detection method?}
An interesting outcome of the observations of bow shocks around exoplanets is that it permits one to infer the magnetic field intensity of the transiting planet. By measuring the difference between the phases at which the near-UV and the optical transits begin ($\Delta \varphi$ illustrated in Figure~\ref{fig.sketch2}), one can derive the stand-off distance from the shock to the centre of the planet, which is assumed to trace the extent of the planetary magnetosphere $r_M$. At the magnetopause, pressure balance between the coronal total pressure and the planet total pressure requires that
\begin{equation}\label{eq.equilibrium}
{\rho_c \Delta u^2} + \frac{[B_c(a)]^2}{8\pi} + p_c= \frac{[B_{p}(r_M)]^2}{8\pi} + p_{p} ,
\end{equation}
where $\rho_c$, $p_c$ and $B_c(a)$ are the local coronal mass density, thermal pressure, and magnetic field intensity at orbital radius $a$, and $p_p$ and $B_p (r_M)$ are the planet thermal pressure and magnetic field intensity at $r_M$. The relative velocity between the material surrounding the planet and the planet itself is $\Delta u$. 
In the case of a magnetised planet with a magnetosphere of a few planetary radii, the planet total pressure is usually dominated by the contribution from the planetary magnetic pressure (i.e., $p_p \sim 0$). \cite{2010ApJ...722L.168V} showed that, in the case of WASP-12b, Eq.~(\ref{eq.equilibrium}) reduces to $B_c(a) \simeq B_p(r_M)$. Further assuming that stellar and planetary magnetic fields are dipolar, we have
\begin{equation}\label{eq.bplanet}
B_p = B_\star \left(  \frac{R_\star/a}{R_{p}/r_M}  \right)^3 ,
\end{equation}
where $B_\star$ and $B_p$ are the magnetic field intensities at the stellar and planetary surfaces, respectively. Eq.~(\ref{eq.bplanet}) shows that the planetary magnetic field can be derived directly from {\it observed} quantities. For WASP-12, using the upper limit of $B_\star<10$~G \citep{2010ApJ...720..872F} and the stand-off distance obtained from the near-UV transit observation $r_M=4.2~R_p$ \citep{2010ApJ...721..923L}, we predicted an upper limit for WASP-12b's planetary magnetic field of $B_p < 24$~G. 

\subsection{Searching for magnetic fields in other exoplanets} \label{sec.othersystems}
In theory, the suggestion that through transit observations one can probe the planetary magnetic field is quite straightforward - all it requires is a measurement of the transit ingress phase in the near-UV. In practice, however, acquisition of near-UV transit data requires the use of space-borne facilities, making follow-ups and new target detections rather difficult. 
 
In order to optimise target selection, \cite{2011MNRAS.411L..46V} presented a classification of the known transiting systems according to their potential for producing shocks that could cause observable light curve asymmetries. The main considered assumption was that, once the conditions for shock formation are met, planetary shocks absorb in certain near-UV lines, in a similar way as WASP-12b. In addition, for it to be detected, the shock must compress the local plasma to a density sufficiently high to cause an observable level of optical depth. This last hypothesis requires the knowledge of the local ambient medium that surrounds the planet. 

By adopting simplified hypotheses, namely that up to the planetary orbit the stellar corona can be treated as in hydrostatic equilibrium and isothermal, \cite{2011MNRAS.411L..46V} predicted the characteristics of the ambient medium that surrounds the planet for a sample of $125$ transiting systems, and discussed whether such characteristics present favourable conditions for the presence and detection of a bow shock. Excluding systems that are quite far ($\gtrsim 400$~pc), the planets that were top ranked are: WASP-19b, WASP-4b, WASP-18b, CoRoT-7b,  HAT-P-7b, CoRoT-1b, TrES-3, and WASP-5b. 

\section{Conclusion}
As the wind outflows from the star, it permeates the interplanetary medium, interacting with any planet encountered on its way. The proper characterisation of stellar winds is therefore crucial to constrain interactions between exoplanets and their surrounding environments and also essential for the study of space weather events on exoplanets. Stellar winds are affect by the stellar rotation, magnetism and coronal temperature, properties that can vary significantly from star to star.  As a consequence, stellar winds of cool stars can actually be significantly different from the solar wind. In this talk, I illustrated how one can take an extra step towards more realistic models of stellar winds of low-mass stars by incorporating observationally reconstructed surface magnetic maps into simulations of stellar winds. I also showed that dramatic differences in stellar magnetism and orbital radius can make the interplanetary medium of exoplanetary systems remarkably distinct from the one present in the solar system. In addition, I showed that the interaction of the stellar winds with exoplanets can lead to observable signatures that are absent in our own solar system. 

\section*{Acknowledgements}
AAV acknowledges support from a Royal Astronomical Society Fellowship and thanks the IAU for a travel support grant to attend the Symposium.

\bibpunct{(}{)}{,}{a}{}{;} 

\def\aj{{AJ}}                   
\def\araa{{ARA\&A}}             
\def\apj{{ApJ}}                 
\def\apjl{{ApJ}}                
\def\apjs{{ApJS}}               
\def\ao{{Appl.~Opt.}}           
\def\apss{{Ap\&SS}}             
\def\aap{{A\&A}}                
\def\aapr{{A\&A~Rev.}}          
\def\aaps{{A\&AS}}              
\def\azh{{AZh}}                 
\def\baas{{BAAS}}               
\def\jrasc{{JRASC}}             
\def\memras{{MmRAS}}            
\def\mnras{{MNRAS}}             
\def\pra{{Phys.~Rev.~A}}        
\def\prb{{Phys.~Rev.~B}}        
\def\prc{{Phys.~Rev.~C}}        
\def\prd{{Phys.~Rev.~D}}        
\def\pre{{Phys.~Rev.~E}}        
\def\prl{{Phys.~Rev.~Lett.}}    
\def\pasp{{PASP}}               
\def\pasj{{PASJ}}               
\def\qjras{{QJRAS}}             
\def\skytel{{S\&T}}             
\def\solphys{{Sol.~Phys.}}      
\def\sovast{{Soviet~Ast.}}      
\def\ssr{{Space~Sci.~Rev.}}     
\def\zap{{ZAp}}                 
\def\nat{{Nature}}              
\def\iaucirc{{IAU~Circ.}}       
\def\aplett{{Astrophys.~Lett.}} 
\def\apspr{{Astrophys.~Space~Phys.~Res.}}   
\def\bain{{Bull.~Astron.~Inst.~Netherlands}}    
\def\fcp{{Fund.~Cosmic~Phys.}}  
\def\gca{{Geochim.~Cosmochim.~Acta}}        
\def\grl{{Geophys.~Res.~Lett.}} 
\def\jcp{{J.~Chem.~Phys.}}      
\def\jgr{{J.~Geophys.~Res.}}    
\def\jqsrt{{J.~Quant.~Spec.~Radiat.~Transf.}}   
\def\memsai{{Mem.~Soc.~Astron.~Italiana}}   
\def\nphysa{{Nucl.~Phys.~A}}    
\def\physrep{{Phys.~Rep.}}      
\def\physscr{{Phys.~Scr}}       
\def\planss{{Planet.~Space~Sci.}}           
\def\procspie{{Proc.~SPIE}}     

\let\astap=\aap
\let\apjlett=\apjl
\let\apjsupp=\apjs
\let\applopt=\ao
\let\mnrasl=\mnras


\begin{thebibliography}{}

\bibitem[\protect\astroncite{{Bouvier} et~al.}{1997}]{1997A&A...326.1023B}
{Bouvier}, J., {Forestini}, M., and {Allain}, S.: 1997,
\newblock {\em \aap} {\bf 326}, 1023

\bibitem[\protect\astroncite{{Catala} et~al.}{2007}]{2007MNRAS.374L..42C}
{Catala}, C., {Donati}, J.-F., {Shkolnik}, E., {Bohlender}, D., and {Alecian},
  E.: 2007,
\newblock {\em \mnrasl} {\bf 374}, L42

\bibitem[\protect\astroncite{{Cranmer}}{2008}]{2008ApJ...689..316C}
{Cranmer}, S.~R.: 2008,
\newblock {\em \apj} {\bf 689}, 316

\bibitem[\protect\astroncite{{Cranmer} and {Saar}}{2011}]{2011ApJ...741...54C}
{Cranmer}, S.~R. and {Saar}, S.~H.: 2011,
\newblock {\em \apj} {\bf 741}, 54

\bibitem[\protect\astroncite{{Donati} and
  {Landstreet}}{2009}]{2009ARA&A..47..333D}
{Donati}, J. and {Landstreet}, J.~D.: 2009,
\newblock {\em \araa} {\bf 47}, 333

\bibitem[\protect\astroncite{{Donati} et~al.}{2008a}]{2008MNRAS.390..545D}
{Donati}, J., {Morin}, J., {Petit}, P., {Delfosse}, X., {Forveille}, T.,
  {Auri{\`e}re}, M., {Cabanac}, R., {Dintrans}, B., {Fares}, R., {Gastine}, T.,
  {Jardine}, M.~M., {Ligni{\`e}res}, F., {Paletou}, F., {Velez}, J.~C.~R., and
  {Th{\'e}ado}, S.: 2008a,
\newblock {\em \mnras} {\bf 390}, 545

\bibitem[\protect\astroncite{{Donati} and {Brown}}{1997}]{1997A&A...326.1135D}
{Donati}, J.-F. and {Brown}, S.~F.: 1997,
\newblock {\em \aap} {\bf 326}, 1135

\bibitem[\protect\astroncite{{Donati} et~al.}{2008b}]{2008MNRAS.385.1179D}
{Donati}, J.-F., {Moutou}, C., {Far{\`e}s}, R., {Bohlender}, D., {Catala}, C.,
  {Deleuil}, M., {Shkolnik}, E., {Collier Cameron}, A., {Jardine}, M.~M., and
  {Walker}, G.~A.~H.: 2008b,
\newblock {\em \mnras} {\bf 385}, 1179

\bibitem[\protect\astroncite{{Falceta-Gon{\c c}alves}
  et~al.}{2006}]{2006MNRAS.368.1145F}
{Falceta-Gon{\c c}alves}, D., {Vidotto}, A.~A., and {Jatenco-Pereira}, V.:
  2006,
\newblock {\em \mnras} {\bf 368}, 1145

\bibitem[\protect\astroncite{{Fares} et~al.}{2009}]{2009MNRAS.398.1383F}
{Fares}, R., {Donati}, J., {Moutou}, C., {Bohlender}, D., {Catala}, C.,
  {Deleuil}, M., {Shkolnik}, E., {Cameron}, A.~C., {Jardine}, M.~M., and
  {Walker}, G.~A.~H.: 2009,
\newblock {\em \mnras} {\bf 398}, 1383

\bibitem[\protect\astroncite{{Fares} et~al.}{2013}]{2013arXiv1307.6091F}
{Fares}, R., {Moutou}, C., {Donati}, J.-F., {Catala}, C., {Shkolnik}, E.,
  {Jardine}, M., {Cameron}, A., and {Deleuil}, M.: 2013,
\newblock {\em \mnras} in press (arXiv:1307.6091)

\bibitem[\protect\astroncite{{Fossati} et~al.}{2010a}]{2010ApJ...720..872F}
{Fossati}, L., {Bagnulo}, S., {Elmasli}, A., {Haswell}, C.~A., {Holmes}, S.,
  {Kochukhov}, O., {Shkolnik}, E.~L., {Shulyak}, D.~V., {Bohlender}, D.,
  {Albayrak}, B., {Froning}, C., and {Hebb}, L.: 2010a,
\newblock {\em \apj} {\bf 720}, 872

\bibitem[\protect\astroncite{{Fossati} et~al.}{2010b}]{2010ApJ...714L.222F}
{Fossati}, L., {Haswell}, C.~A., {Froning}, C.~S., {Hebb}, L., {Holmes}, S.,
  {Kolb}, U., {Helling}, C., {Carter}, A., {Wheatley}, P., {Cameron}, A.~C.,
  {Loeillet}, B., {Pollacco}, D., {Street}, R., {Stempels}, H.~C., {Simpson},
  E., {Udry}, S., {Joshi}, Y.~C., {West}, R.~G., {Skillen}, I., and {Wilson},
  D.: 2010b,
\newblock {\em \apjl} {\bf 714}, L222

\bibitem[\protect\astroncite{{Guedel}}{2004}]{2004A&ARv..12...71G}
{Guedel}, M.: 2004,
\newblock {\em \aapr} {\bf 12}, 71

\bibitem[\protect\astroncite{{Hartmann} and
  {MacGregor}}{1980}]{1980ApJ...242..260H}
{Hartmann}, L. and {MacGregor}, K.~B.: 1980,
\newblock {\em \apj} {\bf 242}, 260

\bibitem[\protect\astroncite{{Hollweg}}{1973}]{1973ApJ...181..547H}
{Hollweg}, J.~V.: 1973,
\newblock {\em \apj} {\bf 181}, 547

\bibitem[\protect\astroncite{{Holzer} et~al.}{1983}]{1983ApJ...275..808H}
{Holzer}, T.~E., {Fla}, T., and {Leer}, E.: 1983,
\newblock {\em \apj} {\bf 275}, 808

\bibitem[\protect\astroncite{{Jardine} et~al.}{2013}]{2013MNRAS.431..528J}
{Jardine}, M., {Vidotto}, A.~A., {van Ballegooijen}, A., {Donati}, J.-F.,
  {Morin}, J., {Fares}, R., and {Gombosi}, T.~I.: 2013,
\newblock {\em \mnras} {\bf 431}, 528

\bibitem[\protect\astroncite{{Jatenco-Pereira} and
  {Opher}}{1989}]{1989A&A...209..327J}
{Jatenco-Pereira}, V. and {Opher}, R.: 1989,
\newblock {\em \aap} {\bf 209}, 327

\bibitem[\protect\astroncite{{Jones} et~al.}{1998}]{1998GeoRL..25.3109J}
{Jones}, G.~H., {Balogh}, A., and {Forsyth}, R.~J.: 1998,
\newblock {\em \grl} {\bf 25}, 3109

\bibitem[\protect\astroncite{{Keppens} and
  {Goedbloed}}{2000}]{2000ApJ...530.1036K}
{Keppens}, R. and {Goedbloed}, J.~P.: 2000,
\newblock {\em \apj} {\bf 530}, 1036

\bibitem[\protect\astroncite{{Lai} et~al.}{2010}]{2010ApJ...721..923L}
{Lai}, D., {Helling}, C., and {van den Heuvel}, E.~P.~J.: 2010,
\newblock {\em \apj} {\bf 721}, 923

\bibitem[\protect\astroncite{{Lazio} et~al.}{2010}]{2010AJ....139...96L}
{Lazio}, T.~J.~W., {Carmichael}, S., {Clark}, J., {Elkins}, E., {Gudmundsen},
  P., {Mott}, Z., {Szwajkowski}, M., and {Hennig}, L.~A.: 2010,
\newblock {\em \aj} {\bf 139}, 96

\bibitem[\protect\astroncite{{Lecavelier des Etangs}
  et~al.}{2013}]{2013A&A...552A..65L}
{Lecavelier des Etangs}, A., {Sirothia}, S.~K., {Gopal-Krishna}, and {Zarka},
  P.: 2013,
\newblock {\em \aap} {\bf 552}, A65

\bibitem[\protect\astroncite{{Lim} and {White}}{1996}]{1996ApJ...462L..91L}
{Lim}, J. and {White}, S.~M.: 1996,
\newblock {\em \apjl} {\bf 462}, L91

\bibitem[\protect\astroncite{{Lima} et~al.}{2001}]{2001A&A...371..240L}
{Lima}, J.~J.~G., {Priest}, E.~R., and {Tsinganos}, K.: 2001,
\newblock {\em \aap} {\bf 371}, 240

\bibitem[\protect\astroncite{{Llama} et~al.}{2013}]{2013arXiv1309.2938L}
{Llama}, J., {Vidotto}, A.~A., {Jardine}, M., {Wood}, K., {Fares}, R., and
  {Gombosi}, T.~I.: 2013, \newblock {\em \mnrasl}, in press (arXiv: 1309.2938)

\bibitem[\protect\astroncite{{Llama} et~al.}{2011}]{2011MNRAS.416L..41L}
{Llama}, J., {Wood}, K., {Jardine}, M., {Vidotto}, A.~A., {Helling}, C.,
  {Fossati}, L., and {Haswell}, C.~A.: 2011,
\newblock {\em \mnrasl} {\bf 416}, L41

\bibitem[\protect\astroncite{{McComas} et~al.}{1995}]{1995JGR...10019893M}
{McComas}, D.~J., {Barraclough}, B.~L., {Gosling}, J.~T., {Hammond}, C.~M.,
  {Phillips}, J.~L., {Neugebauer}, M., {Balogh}, A., and {Forsyth}, R.~J.:
  1995,
\newblock {\em \jgr} {\bf 100}, 19893

\bibitem[\protect\astroncite{{McComas} et~al.}{2008}]{2008GeoRL..3518103M}
{McComas}, D.~J., {Ebert}, R.~W., {Elliott}, H.~A., {Goldstein}, B.~E.,
  {Gosling}, J.~T., {Schwadron}, N.~A., and {Skoug}, R.~M.: 2008,
\newblock {\em \grl} {\bf 35}, 18103

\bibitem[\protect\astroncite{{Mestel}}{1968}]{1968MNRAS.138..359M}
{Mestel}, L.: 1968,
\newblock {\em \mnras} {\bf 138}, 359

\bibitem[\protect\astroncite{{Morin} et~al.}{2008}]{2008MNRAS.390..567M}
{Morin}, J., {Donati}, J., {Petit}, P., {Delfosse}, X., {Forveille}, T.,
  {Albert}, L., {Auri{\`e}re}, M., {Cabanac}, R., {Dintrans}, B., {Fares}, R.,
  {Gastine}, T., {Jardine}, M.~M., {Ligni{\`e}res}, F., {Paletou}, F., {Ramirez
  Velez}, J.~C., and {Th{\'e}ado}, S.: 2008,
\newblock {\em \mnras} {\bf 390}, 567

\bibitem[\protect\astroncite{{Parker}}{1958}]{1958ApJ...128..664P}
{Parker}, E.~N.: 1958,
\newblock {\em \apj} {\bf 128}, 664

\bibitem[\protect\astroncite{{Petit} et~al.}{2008}]{2008MNRAS.388...80P}
{Petit}, P., {Dintrans}, B., {Solanki}, S.~K., {Donati}, J.-F., {Auri{\`e}re},
  M., {Ligni{\`e}res}, F., {Morin}, J., {Paletou}, F., {Ramirez Velez}, J.,
  {Catala}, C., and {Fares}, R.: 2008,
\newblock {\em \mnras} {\bf 388}, 80

\bibitem[\protect\astroncite{{Pinto} et~al.}{2011}]{2011ApJ...737...72P}
{Pinto}, R.~F., {Brun}, A.~S., {Jouve}, L., and {Grappin}, R.: 2011,
\newblock {\em \apj} {\bf 737}, 72

\bibitem[\protect\astroncite{{Pneuman} and {Kopp}}{1971}]{1971SoPh...18..258P}
{Pneuman}, G.~W. and {Kopp}, R.~A.: 1971,
\newblock {\em \solphys} {\bf 18}, 258

\bibitem[\protect\astroncite{{Smith} et~al.}{2009}]{2009MNRAS.395..335S}
{Smith}, A.~M.~S., {Collier Cameron}, A., {Greaves}, J., {Jardine}, M.,
  {Langston}, G., and {Backer}, D.: 2009,
\newblock {\em \mnras} {\bf 395}, 335

\bibitem[\protect\astroncite{{Suess} and {Smith}}{1996}]{1996GeoRL..23.3267S}
{Suess}, S.~T. and {Smith}, E.~J.: 1996,
\newblock {\em \grl} {\bf 23}, 3267

\bibitem[\protect\astroncite{{Suzuki} et~al.}{2012}]{2012arXiv1212.6713S}
{Suzuki}, T.~K., {Imada}, S., {Kataoka}, R., {Kato}, Y., {Matsumoto}, T.,
  {Miyahara}, H., and {Tsuneta}, S.: 2012, PASJ in press (arXiv:1212.6713)
  
\bibitem[\protect\astroncite{{Tsinganos} and {Low}}{1989}]{1989ApJ...342.1028T}
{Tsinganos}, K. and {Low}, B.~C.: 1989,
\newblock {\em \apj} {\bf 342}, 1028

\bibitem[\protect\astroncite{{Vidotto}}{2013}]{2013A&G....54a1.25V}
{Vidotto}, A.: 2013,
\newblock {\em Astronomy and Geophysics} {\bf 54(1)}, 010001

\bibitem[\protect\astroncite{{Vidotto} et~al.}{2012}]{2012MNRAS.423.3285V}
{Vidotto}, A.~A., {Fares}, R., {Jardine}, M., {Donati}, J.-F., {Opher}, M.,
  {Moutou}, C., {Catala}, C., and {Gombosi}, T.~I.: 2012,
\newblock {\em \mnras} {\bf 423}, 3285

\bibitem[\protect\astroncite{{Vidotto} et~al.}{2010a}]{2010ApJ...722L.168V}
{Vidotto}, A.~A., {Jardine}, M., and {Helling}, C.: 2010a,
\newblock {\em \apjl} {\bf 722}, L168

\bibitem[\protect\astroncite{{Vidotto} et~al.}{2011a}]{2011MNRAS.411L..46V}
{Vidotto}, A.~A., {Jardine}, M., and {Helling}, C.: 2011a,
\newblock {\em \mnrasl} {\bf 411}, L46

\bibitem[\protect\astroncite{{Vidotto} et~al.}{2011b}]{2011MNRAS.414.1573V}
{Vidotto}, A.~A., {Jardine}, M., and {Helling}, C.: 2011b,
\newblock {\em \mnras} {\bf 414}, 1573

\bibitem[\protect\astroncite{{Vidotto} and
  {Jatenco-Pereira}}{2006}]{2006ApJ...639..416V}
{Vidotto}, A.~A. and {Jatenco-Pereira}, V.: 2006,
\newblock {\em \apj} {\bf 639}, 416

\bibitem[\protect\astroncite{{Vidotto} et~al.}{2009a}]{2009ApJ...703.1734V}
{Vidotto}, A.~A., {Opher}, M., {Jatenco-Pereira}, V., and {Gombosi}, T.~I.:
  2009a,
\newblock {\em \apj} {\bf 703}, 1734

\bibitem[\protect\astroncite{{Vidotto} et~al.}{2009b}]{2009ApJ...699..441V}
{Vidotto}, A.~A., {Opher}, M., {Jatenco-Pereira}, V., and {Gombosi}, T.~I.:
  2009b,
\newblock {\em \apj} {\bf 699}, 441

\bibitem[\protect\astroncite{{Vidotto} et~al.}{2010b}]{2010ApJ...720.1262V}
{Vidotto}, A.~A., {Opher}, M., {Jatenco-Pereira}, V., and {Gombosi}, T.~I.:
  2010b,
\newblock {\em \apj} {\bf 720}, 1262

\bibitem[\protect\astroncite{{Wargelin} and
  {Drake}}{2002}]{2002ApJ...578..503W}
{Wargelin}, B.~J. and {Drake}, J.~J.: 2002,
\newblock {\em \apj} {\bf 578}, 503

\bibitem[\protect\astroncite{{Washimi} and
  {Shibata}}{1993}]{1993MNRAS.262..936W}
{Washimi}, H. and {Shibata}, S.: 1993,
\newblock {\em \mnras} {\bf 262}, 936

\bibitem[\protect\astroncite{{Wilhelm}}{2006}]{2006A&A...455..697W}
{Wilhelm}, K.: 2006,
\newblock {\em \aap} {\bf 455}, 697

\bibitem[\protect\astroncite{{Wood} et~al.}{2001}]{2001ApJ...547L..49W}
{Wood}, B.~E., {Linsky}, J.~L., {M{\"u}ller}, H., and {Zank}, G.~P.: 2001,
\newblock {\em \apjl} {\bf 547}, L49

\bibitem[\protect\astroncite{{Zarka}}{2007}]{2007P&SS...55..598Z}
{Zarka}, P.: 2007,
\newblock {\em \planss} {\bf 55}, 598

\end{thebibliography}
\end{document}